\documentclass[floats,floatfix,showpacs,amssymb,prd,nofootinbib,superscriptaddress,aps,notitlepage,reprint]{revtex4-1}
\usepackage[utf8,latin1]{inputenc}
\usepackage[english]{babel}
\usepackage{graphicx, epsfig, amssymb} %include figure files
\usepackage{amsmath, amsfonts}
\usepackage{epstopdf}
\usepackage{bm} %include bold math: \bm{} creates bold letters in math mode
\usepackage[normalem]{ulem} % either use this (simple) or
\usepackage{soul}
\usepackage[linktocpage]{hyperref}
\usepackage[caption=false]{subfig}
\usepackage[usenames]{color}
\usepackage{lipsum}
\def\L{\mathcal{L}}
\def\p{\partial}

\newcommand{\diff}{\mathrm{d}}
\newcommand{\beq}{\begin{equation}}
\newcommand{\eeq}{\end{equation}}
\newcommand{\ben}{\begin{enumerate}}
\newcommand{\een}{\end{enumerate}}
\newcommand{\bi}{\begin{itemize}}
\newcommand{\ei}{\end{itemize}}

\def\ga{\mathrel{\raise.3ex\hbox{$>$\kern-.75em\lower1ex\hbox{$\sim$}}}}
\def\la{\mathrel{\raise.3ex\hbox{$<$\kern-.75em\lower1ex\hbox{$\sim$}}}}

\newcommand{\beqal}{\begin{eqnarray}\label}
\newcommand{\none}{\end{eqnarray}}
\newcommand{\beqa}{\begin{eqnarray}}
\newcommand{\eeqa}{\end{eqnarray}}

\begin{document}
\title{\large Scalar radiation from a source rotating around a regular black hole}

\author{Rafael P. Bernar}\email{rafael.bernar@pq.cnpq.br}
\affiliation{Faculdade de F\'{\i}sica, Universidade Federal do Par\'a, 66075-110, Bel\'em, Par\'a, Brazil.}
\affiliation{Departamento de F\'isica, Universidade Federal do Maranh\~ao, Campus Universit\'ario do Bacanga, 65080-805, S\~ao Lu\'is, Maranh\~ao, Brazil.}

\author{Lu\'is C. B. Crispino}\email{crispino@ufpa.br}
\affiliation{Faculdade de F\'{\i}sica, Universidade Federal do Par\'a, 66075-110, Bel\'em, Par\'a, Brazil.}
	
\begin{abstract}	
The scalar radiation emitted by a source in geodesic circular orbit around a regular Bardeen black hole is analyzed. We use the quantum field theory in curved spacetime framework to obtain the emitted power of radiation by computing the one-particle emission amplitude of scalar particles in the curved background. We compare our results to a similar setting in the spacetime of a Reissner-Nordstr\"{o}m black hole.
\end{abstract}
	
\pacs{
%	04.60.-m, %%Quantum gravity
	04.62.+v, %%Quantum fields in curved spacetime
%	04.50.-h, %%Higher dimensional gravity
	04.25.Nx, %%Perturbation theory
%	04.60.Gw, %%Covariant quantization
	11.25.Db  %%Properties of perturbation theory
	}
	
\date{\today}
	
\maketitle

%\tableofcontents

%%%%%%%%%%%%%%%%INTRODUCTION%%%%%%%%%%%%%%%%%%
\section{Introduction}
%%%%%%%%%%%%%%%%%%%%%%%%%%%%%%%%%%%%%%%%%%%%%%

Black holes (BHs) play a central part in General Relativity (GR) and other theories of gravity. Their nontrivial causal structure, particularly the presence of an event horizon, provides an interesting setting to test fundamental physics. Moreover, these properties of BH spacetimes, and their strong gravity regime, are interesting in their own right and may have implications for the very nature of spacetime. In particular, BH solutions are usually plagued with singularities, where GR, as a description of spacetime physics, breaks down.

%One important prospect of modifications of GR is to 
Within GR, the occurrence of singularities is quite generic, in the sense that the singularity theorems impose severe constraints on the types of matter which can avoid them~\cite{Hawking1975a}. Additionally, it is expected that almost all singularities are protected by event horizons, according to the weak cosmic censorship conjecture~\cite{Penrose1969,Wald1997}. Hence, usually, a singularity is expected to be located in the inside region of a BH. There are, however, BH solutions with no singularities, the so-called regular BHs. The first regular BH solution was presented in Ref.~\cite{Bardeen1968} (See Ref.~\cite{Ansoldi2008} for an extensive review about regular BHs). Although most, if not all, regular BH solutions include exotic matter, it is also a possibility that common matter in alternative theories of gravity gives rise to entirely regular solutions. This may be the case if regular BHs solutions are viewed as an effective description to spacetime in a quantum theory of gravity~\cite{Carballo-Rubio2018}.

The interaction of BHs with fundamental test fields is important, not only for its theoretical implications, but also because it can be useful in the investigation of the BH properties, through the imprints left by the geometry on the observables. It is the case, for example, of the BH quasinormal modes, which are instantiated by the test field modes~\cite{Nollert1999,Berti2009,Konoplya2011}. Geometry signatures may also be found in scattering and absorption data. The analysis of emitted radiation by matter surrounding a BH is particularly important to infer properties of both matter and spacetime. In this context, radiation settings in BHs spacetimes have been extensively studied since the 1970s. Moreover, the recent detections of gravitational waves and electromagnetic counterparts in a binary neutron star inspiral (see Ref.~\cite{LIGOScientificCollaborationandVirgoCollaboration2017} and references therein), have drawn additional attention to this research subject.

The gravitational radiation emitted by a source that is falling radially into a Schwarzschild BH was computed in Refs.~\cite{Davis1971,Davis1971a}. Due to its simplicity, a massless scalar field can be used as a model to study more general (electromagnetic or gravitational) radiation emission processes by sources near BHs. For the source orbiting the BH in a circular geodesic orbit, the so-called geodesic synchrotron radiation setting, 
the massless scalar field model was used to compute the high-frequency radiation emission as a first approximation to the gravitational radiation case in Ref.~\cite{Breuer1973}. Using the quantum field theory in curved spacetimes framework, the scalar radiation emission in Schwarzschild spacetime was revisited in Refs.~\cite{Crispino2000,*Crispino2016,Castineiras2007,Crispino2008, Oliveira2018}. In the context of geodesic synchrotron radiation, this framework has been used to analyze the emission of gravitational radiation in Ref.~\cite{Bernar2017}. There are also other results for emitted radiation in BH spacetimes such as the electromagnetic field in a Schwarzschild BH~\cite{Castineiras2005} and the massless scalar field in Kerr BH~\cite{Macedo2012}. Regarding regular BHs, absorption and scattering of scalar waves by Bardeen BHs were studied in Refs.~\cite{Macedo2014,Macedo2015,Macedo2016}.

%As was previously mentioned, the radiation emitted by matter surrounding a regular BH can be used to analyze some features of the spacetime.
In this paper, we consider the emission of scalar radiation by a source in circular motion around a Bardeen BH. We compare our results to the case of a source orbiting a Reissner-Nordstr\"{o}m BH. This paper is organized as follows. In Sec.~\ref{sec:bardeenbh}, we review the spacetime of a Bardeen BH and the corresponding circular geodesics. In Sec.~\ref{sec:scalar-radiation} we revisit the framework of a quantum scalar field theory in the curved spacetime of a spherically symmetric BH. We apply this framework to the computation of the emitted power of scalar radiation by a source in circular geodesic orbit around the BH. We present our numerical results in Sec.~\ref{sec:results}, comparing them with the corresponding ones for Reissner-Nordstr\"{o}m BHs. We end this paper with some remarks in Sec.~\ref{sec:finalremarks}. We use natural units such that $c=G=\hbar=1$.

%%%%%%%%%%%%%%%%%%%MAIN%%%%%%%%%%%%%%%%%%%%%%%

%%%%%%%%%%%%%%%%%%%%%%%%%%%%%%%%%%%%%%%%%%%%%%
\section{Bardeen regular black holes}
\label{sec:bardeenbh}
%%%%%%%%%%%%%%%%%%%%%%%%%%%%%%%%%%%%%%%%%%%%%%

The line element of a spherically symmetric BH spacetime can be written as
\beq
\diff s^2=-f(r) \diff t^2+\frac{\diff r}{g(r)}+r^2(\diff \theta^2+\sin \theta^2 \diff \phi^2). \label{eq:sphericalmetric}
\eeq
For the so-called Bardeen black hole, we have 
\beq
f(r)=g(r)=1-\frac{2 M r^2}{(r^2+q_B^2)^{\frac{3}{2}}},
\eeq
where the parameter $M$ is associated to the mass of the BH. The parameter $q_B$ was identified in Ref.~\cite{Ayon-Beato2000} as the charge of a magnetic monopole in a theory of nonlinear electrodynamics, described by the following action
\beq
S=S_{EH}+\int \diff^4 x \sqrt{-g} \frac{\L(F)}{4\pi},
\eeq
where $S_{EH}$ is the Einstein-Hilbert action. The Lagrangian $\L(F)$ associated to the electromagnetic field strength $F_{ab}$, with $F=\frac{1}{4}F_{ab}F^{ab}$, is given by
\beq
\L(F)=\frac{3}{2sq_B^2}\left(\frac{\sqrt{2 q_B^2 F}}{1-\sqrt{2 q_B^2 F}}\right)^{5/2}
\eeq
where $s=|q_B|/2M$. 

The existence of zeroes of the function $f(r)$ depends on the parameters $M$ and $q_B$. If $0 \leq q_B < q_B^{\mathrm{ext}}$, where $q_B^{\mathrm{ext}} \equiv 4 M/(3 \sqrt{3})$, $f(r)$ vanishes at two values of $r$, $r=r_{-}$ and $r=r_{+}$, which are associated to two horizons. One of them ($r=r_{-}$) is a Cauchy horizon and the other ($r=r_{+}$) is an event horizon. For $q=q_B^{\mathrm{ext}}$, the two horizons coincide. For $q > q_B^{\mathrm{ext}}$ we have no horizons. We shall consider only the cases in which $0 \leq q \leq q_B^{\mathrm{ext}}$. 

The causal structure of the Bardeen BH is very similar to the causal structure of an electrically charged BH, the so-called Reissner-Nordstr\"{o}m (RN) BH, which is described by the line element (\ref{eq:sphericalmetric}) with
\beq
f(r)=g(r)=1-\frac{2M}{r}+\frac{q_{RN}^2}{r^2}. \label{eq:RN-function}
\eeq
In Eq.~(\ref{eq:RN-function}), $q_{RN}$ is the electric charge of the BH. The RN solution also presents two horizons for $0 \leq q_{RN} < q_{RN}^{\mathrm{ext}}$, where $q_{RN}^{\mathrm{ext}} \equiv M$. In the extreme case ($q_{RN}=q_{RN}^{\mathrm{ext}}$), the two horizons coincide. For $q > q_{RN}^{\mathrm{ext}}$, we have a naked singularity.

We aim to compare the emitted power of scalar radiation in Bardeen and RN spacetimes. For the RN case, this was computed in Ref.~\cite{Crispino2009}. Although the charges in the RN and Bardeen spacetimes are associated to different fields, we expect a somewhat similar behavior for the emission of scalar radiation by a source in circular orbit in these two spacetimes. In order to better compare the two situations, we parametrize our results by the normalized charge, namely
\beq
Q_{(i)} \equiv q_{(i)}/q_{(i)}^{\mathrm{ext}},
\eeq
 with $(i)=B,RN$ and there is no implicit sum in the $(i)$ subscripts. 
\subsection{Circular geodesics in a Bardeen BH}

Let us briefly review the circular geodesic orbits in a spherically symmetric spacetime such as the Bardeen BH spacetime. Following Refs.~\cite{Chandrasekhar1983,Cardoso2009}, the equations of motion governing geodesics can be derived from the following Lagrangian
\beq
2\L = g_{\mu\nu}\dot{x}^{\mu}\dot{x}^{\nu},
\eeq
where the overdot denotes differentiation with respect to an affine parameter. (For timelike geodesics, the particle's proper time can be considered as the affine parameter.) For spacetimes described by Eq.~(\ref{eq:sphericalmetric}), with $f(r)=g(r)$, the Lagrangian can be written as
\beq
2\L = -f(r)\dot{t}^2+\frac{\dot{r}^2}{f(r)}+r^2\dot{\theta}^2+r^2\sin^2\theta \ \dot{\phi}^2.
\eeq
Since this Lagrangian is independent of both $t$ and $\phi$, it follows from the Euler-Lagrange equations for these coordinates that the canonical momenta, given by
\beqa
p_t&=&-\frac{\p \L}{\p \dot{t}}=f(r)\dot{t}=E, \\
p_{\phi}&=& \frac{\p \L}{\p \dot{\phi}}=r^2\dot{\phi}=L,
\eeqa
are integrals of motion. 
Considering the motion in the equatorial plane ($\theta=\pi/2$, $\dot{\theta}=0$) and noting that $2\L = -1$ for timelike geodesics, the equation of motion for the particle's radial coordinate may be written as
\beq
\dot{r}^2=f(r)\left[E^2-\frac{L^2}{r^2}-1\right].
\eeq
For a circular orbit, with $r=R$, the $\dot{r}=\ddot{r}=0$ equations yield
\beq
E^2=\frac{2f(R)}{2f(R)-Rf'(R)}, \ \ \ L^2=\frac{R^3f'(R)}{2f(R)-Rf'(R)},
\eeq
which, since $E^2$ and $L^2$ are positive quantities, imply
\beq
2f(R)-Rf'(R)>0. \label{eq:timelike-inequality}
\eeq
Solving the equation
\beq
R_{\mathrm{Light}}=\frac{2f(R_{\mathrm{Light}})}{f'(R_{\mathrm{Light}})}, \label{eq:photonsphere}
\eeq
one obtains the radial position $R_{\mathrm{Light}}$ of the light ring, which is a planar circular null geodesic~\cite{Cardoso2009}. In view of Eq.~(\ref{eq:timelike-inequality}), $R_{\mathrm{Light}}$ is also the limiting value for the radius of the circular timelike geodesics. The orbital angular velocity of the circular geodesics is given by
\beq
\Omega=\frac{\diff \phi}{\diff t}=\frac{\dot{\phi}}{\dot{t}}=\sqrt{\frac{f'(R)}{2R}}=\sqrt{\frac{M(R^2-2q_B^2)}{(R^2+q_B^2)^{5/2}}}, \label{eq:angular-velocity}
\eeq
where the last equality is valid for a Bardeen BH.

%%%%%%%%%%%%%%%%%%%%%%%%%%%%%%%%%%%%%%%%%%%%%%
\section{Scalar radiation and emitted power}
\label{sec:scalar-radiation}
\begin{figure}
	\includegraphics[scale=0.7]{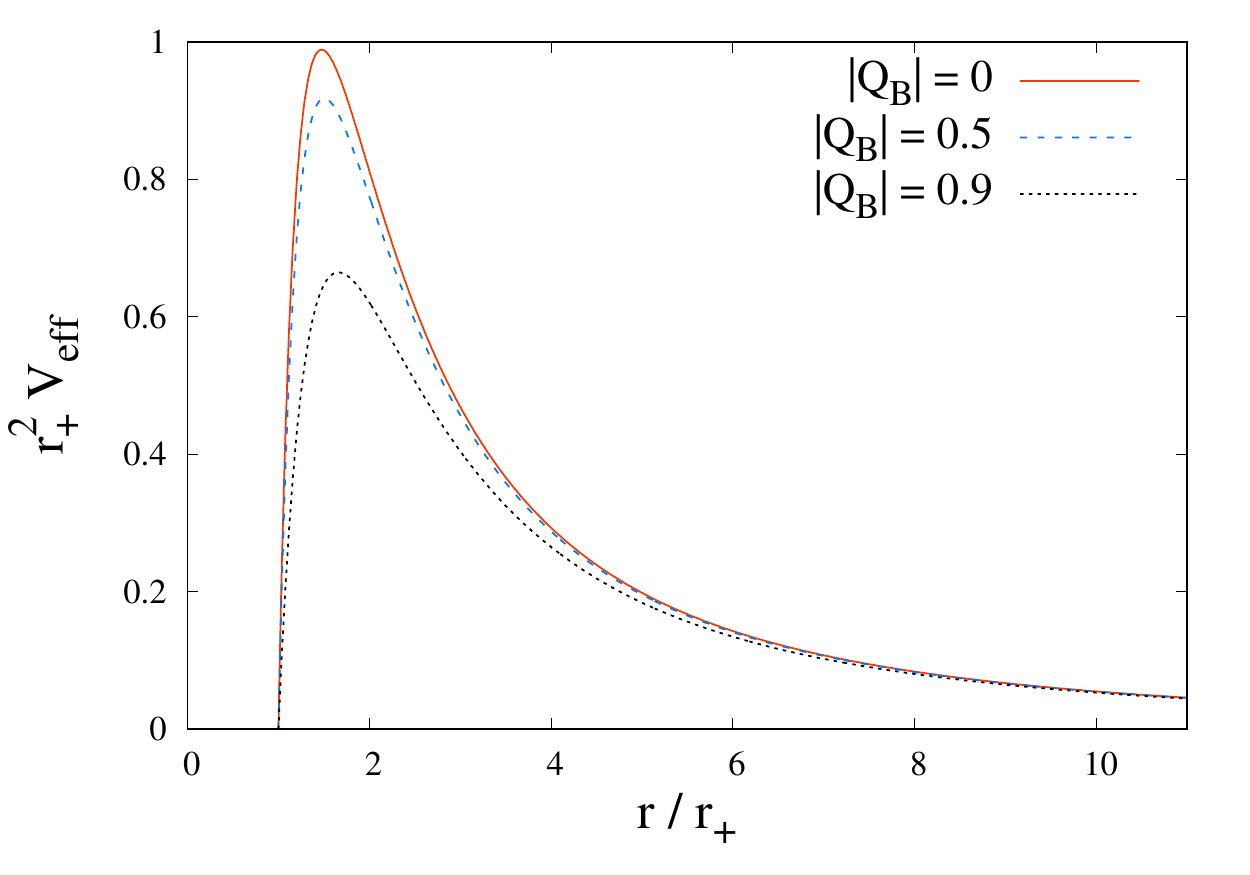}
	\caption{Effective potential, given by Eq.~(\ref{eq:effective-potential}), plotted for $l=1$, as a function of $r/r_{+}$. We compare the Schwarzschild BH case ($|Q_B|=0$) with the $|Q_B|=0.5$, and $|Q_B|=0.9$ Bardeen BH cases.}
	\label{fig:effectivepontential}
\end{figure}

%%%%%%%%%%%%%%%%%%%%%%%%%%%%%%%%%%%%%%%%%%%%%%
The quantization of a massless scalar field in the spacetime outside a Bardeen BH is very similar to the same procedure in the Schwarzschild spacetime. In this Section, we follow mainly Ref.~\cite{Crispino2000,Crispino2016}.
The massless scalar field $\Phi(x)$ obeys the Klein-Gordon equation, namely
\beq
\nabla^{\mu}\nabla_{\mu} \Phi(x) =0. \label{eq:scalarfieldeq}
\eeq
Positive-frequency solutions to Eq.~(\ref{eq:scalarfieldeq}), with respect to the timelike Killing vector field $\partial_t$, can be written as
\beq
u_{\omega l m}(x)=\sqrt{\frac{\omega}{\pi}}\frac{\psi_{\omega l}(r)}{r}Y_{lm}(\theta,\phi)e^{-i \omega t} \ \ \ (\omega > 0), \label{eq:field-decompostion}
\eeq
where $Y_{lm}(\theta,\phi)$ are the scalar spherical harmonics. 
The functions $\psi_{\omega l}(r)$ satisfy the following Schr\"{o}dinger-like equation:
\beq
\left[-f(r)\frac{\diff}{\diff r}\left(f(r)\frac{\diff}{\diff r}\right)+V_{\mathrm{eff}}(r)\right]\psi_{\omega l}(r)=\omega^2\psi_{\omega l}(r), \label{eq:radial-part-eq}
\eeq
where the effective potential is given by (see Fig.~\ref{fig:effectivepontential})
\beq
V_{\mathrm{eff}}(r)=f(r)\left[\frac{l(l+1)}{r^2}+\frac{f'(r)}{r}\right]. \label{eq:effective-potential}
\eeq

The asymptotic forms of the two independent solutions to Eq.~(\ref{eq:radial-part-eq}) are given by
\beq
\psi^{up}_{\omega l}\approx \begin{cases} A^{up}_{\omega l}\left(e^{i\omega r^{*}}+\mathcal{R}^{up}_{\omega l}e^{-i\omega r^{*}}\right), \hspace{0.25cm}r \gtrsim r_{+},\\
	A^{up}_{\omega l}\mathcal{T}^{up}_{\omega l}e^{i\omega r^{*}}, \hspace{0.5cm} r \gg r_{+}; \end{cases} \label{eq:up-boundary-conditions}
\eeq
\beq
\psi^{in}_{\omega l} \approx \begin{cases} A^{in}_{\omega l}\mathcal{T}^{in}_{\omega l}e^{-i\omega r^{*}}, \hspace{0.5cm}r \gtrsim r_{+},\\
	A^{in}_{\omega l}\left(e^{-i \omega r^{*}}+\mathcal{R}^{in}_{\omega l}e^{i \omega r^{*}}\right), \hspace{.25cm} r \gg r_{+}, \end{cases} \label{eq:in-boundary-conditions}
\eeq
where $r^{*}$ is the tortoise coordinate defined by $\diff r^{*} \equiv f^{-1}\diff r$. The $up$ solutions represent modes purely incoming from the past event horizon $\mathcal{H}^{-}$, while the $in$ solutions represent modes purely incoming from the past null infinity $\mathcal{J}^{-}$. 

The quantum field operator $\hat{\Phi}(x)$ is expanded in terms of positive-frequency solutions $u^{P}_{\omega l m}(x)$ to Eq.~(\ref{eq:scalarfieldeq}) and their complex conjugates, namely
\beq
\hat{\Phi}(x)=\sum_{P} \sum_{l,m}\int_{0}^{\infty} \diff \omega \left[ u^{P}_{\omega l m}(x) \hat{a}^{P}_{\omega l m} + \mathrm{H.c.} \right],
\eeq
where the superscript $P$ stands for $in$ and $up$ modes.
We normalize the modes $u^{P}_{\omega l m}(x)$ according to the Klein-Gordon inner product, defined by
\beq
\langle \Phi, \Psi \rangle \equiv i \int_{\Sigma} \diff \Sigma^{\mu} \left(\overline{\Phi}\nabla_{\mu}\Psi-\Psi\nabla_{\mu}\overline{\Phi}\right), \label{eq:innerproduct}
\eeq
where the overbar denotes complex conjugation and $\Sigma$ is a Cauchy hypersurface. It can readily be shown that the inner product (\ref{eq:innerproduct}) is independent of the choice of the hypersurface $\Sigma$, if $\Phi$ and $\Psi$ satisfy the equations of motion (\ref{eq:scalarfieldeq})~\cite{Friedman1978}. By requiring that
\beq
\langle u^{P}_{\omega l m}, u^{P'}_{\omega' l' m'} \rangle=\delta^{PP'}\delta^{ll'}\delta^{mm'}\delta(\omega-\omega'),
\eeq
the creation and annihilation operators, $\hat{a}^{P\dagger}_{\omega l m}$ and $\hat{a}^{P}_{\omega l m}$, satisfy the usual nonvanishing commutation relations, given by
\beq
[\hat{a}^{P}_{\omega l m},\hat{a}^{P'\dagger}_{\omega' l' m'}]=\delta^{PP'}\delta^{ll'}\delta^{mm'}\delta(\omega-\omega').
\eeq

Using the inner product~(\ref{eq:innerproduct}), we obtain the overall normalization constants in Eqs.~(\ref{eq:up-boundary-conditions}) and~(\ref{eq:in-boundary-conditions}), namely
\beq
A^{in}_{\omega l}=A^{up}_{\omega l}=\frac{1}{2 \omega}.
\eeq

We consider a scalar source $j(x)$, in circular geodesic orbit around the Bardeen BH, interacting with the scalar field via the following contribution to the action:
\beq
\hat{S}_I=\int \diff^4 x \sqrt{-g} \ j(x)\hat{\Phi}(x).
\eeq
For the source moving with constant angular velocity $\Omega$ (as measured by asymptotic static observers), at $r=R$, in the plane $\theta=\pi/2$, we may write
\beq
j(x)=\frac{\lambda}{u^{t}\sqrt{-g}}\delta(r-R)\delta(\theta - \pi/2)\delta(\phi - \Omega t), \label{eq:source-current}
\eeq
where $\lambda$ is a small coupling constant. The $4$-velocity of the rotating source is 
\beq
u^{\mu}=(\gamma,0,0,\gamma \Omega),
\eeq
where
\beq
\gamma=\frac{\diff t}{\diff \tau}=\frac{1}{[f(R)-R^2\Omega^2]^{\frac{1}{2}}}.
\eeq

To first order in perturbation theory, the emission amplitude of a $P=in, up$ scalar particle with quantum numbers $l,m$, and frequency $\omega$, is given by
\beq
\mathcal{A}^{P;\omega l m}_{\mathrm{em}}= \langle P; \omega l m| i \hat{S}_I | 0 \rangle = i \int \diff^4 x \sqrt{-g} \ j(x) \overline{u^{P}_{\omega l m}}(x). \label{eq:emissionamplitude}
\eeq 
The vacuum $| 0 \rangle$ is the quantum state annihilated by all the $\hat{a}^{P}_{\omega l m}$, i.e. it is a Boulware-like vacuum~\cite{Boulware1975} for the Bardeen BH spacetime. For the Unruh- or Hartle-Hawking-like vacuum states~\cite{Unruh1976,Hartle1976}, the power associated with Eq.~(\ref{eq:emissionamplitude}) would be the one obtained for the net radiation emitted by the source, since the absorption and stimulated emission rates induced by the thermal fluxes are exactly the same. Due to the source's structure, given by Eq.~(\ref{eq:source-current}), the emission amplitude is proportional to $\delta(\omega - m \Omega)$, i.e. there is only emission of scalar particles with $\omega=m \Omega$.

The emitted power is then given by
\beq
W^{P;lm}_{\mathrm{em}}=\int_{0}^{\infty} \diff \omega \ \omega \ |\mathcal{A}^{P;\omega l m}_{\mathrm{em}}|^2 / T,
\eeq
where $T=\int_{-\infty}^{\infty} \diff t=2 \pi \delta(0)$ is the total time as measured by an asymptotic static observer~\cite{Higuchi1998,Crispino1998}.
By using Eq.~(\ref{eq:field-decompostion}) to compute the emission amplitude given by Eq.~(\ref{eq:emissionamplitude}), we find the emitted power to be
\beqa
W^{P;lm}_{\mathrm{em}}=2 \omega_m^2 \lambda^2 [f(R)-R^2 \Omega^2]\left|\frac{\psi^{P}_{\omega_m l}}{R}\right|^2\left|Y_{lm}(\pi/2,\Omega t)\right|^2, \nonumber \\
\eeqa
where $\omega_m \equiv m \Omega$. The $\left|Y_{lm}(\pi/2,\Omega t)\right|^2$ is a time independent quantity, which is zero for odd values of $l+m$ and 
\beq
\left|Y_{lm}(\pi/2,\Omega t)\right|^2=\frac{2l+1}{4 \pi}\frac{(l+m-1)!!(l-m-1)!!}{(l+m)!!(l-m)!!}
\eeq
for even values of $l+m$~\cite{Gradshteyn1980}.

%%%%%%%%%%%%%%%%%%%%%%%%%%%%%%%%%%%%%%%%%%%%%%
\section{Results}
\label{sec:results}
%%%%%%%%%%%%%%%%%%%%%%%%%%%%%%%%%%%%%%%%%%%%%%

To compute the power of scalar radiation emitted by the source in circular orbit around the Bardeen BH, we have to numerically integrate Eq.~(\ref{eq:radial-part-eq}), in order to obtain the radial functions $\psi^{up}_{\omega l}$ and $\psi^{in}_{\omega l}$. For circular geodesics, we can invert Eq.~(\ref{eq:angular-velocity}) to find $R$ as a function of $\Omega$, so that the emitted power can be written exclusively as a function of quantities measured at infinity, namely $M$ and $\Omega$. We can thus obtain numerically the emitted power for arbitrary circular geodesic orbits, i.e. for different values of the associated angular velocity $\Omega$. For the numerical integration of Eq.~(\ref{eq:radial-part-eq}), we use the same procedure as described in Ref.~\cite{Bernar2017}.

\begin{figure}
	\includegraphics[scale=0.7]{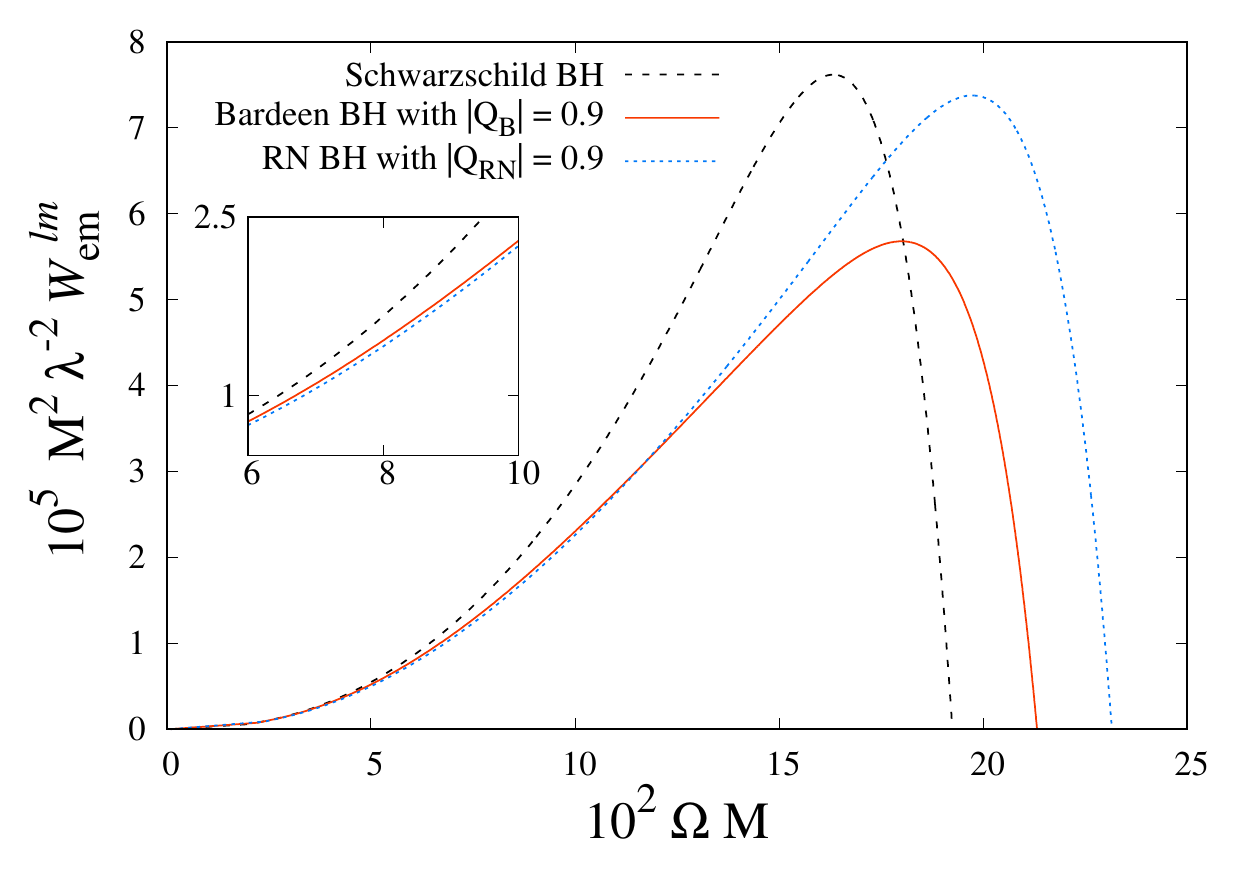}
	\caption{Emitted power of the mode $l=m=1$ as a function of $\Omega$, for the source rotating around the following BH in cases: (i) a Schwarzschild BH; (ii) a Bardeen BH with $|Q_B|=0.9$; (iii) a Reissner-Nordstr\"{o}m BH with $|Q_{RN}|=0.9$.}
	\label{fig:comparison-W}
\end{figure}

In Fig.~\ref{fig:comparison-W}, we compare the emitted power for a given choice of $l$ and $m$, namely
\beq
W^{lm}_{\mathrm{em}}=W^{in;lm}_{\mathrm{em}}+W^{up;lm}_{\mathrm{em}},
\eeq 
 for the mode $l=m=1$, when the source is rotating around a Schwarzschild BH, a Bardeen BH with $|Q_B|=0.9$ or a RN BH with $|Q_{RN}|=0.9$. For low angular velocities, the emitted power in the Bardeen BH case is slightly greater than the one in the RN case. However, as we increase $\Omega$, the RN emitted power starts to dominate, in comparison with the Bardeen one. This happens generically for any mode.
This is in accordance with the fact that, far away from the BH, where $\Omega \to 0$, the charge contribution to the metric falls off as $r^{-3}$ in the Bardeen case, whereas it behaves like $r^{-2}$ in the RN spacetime. Thus, far away from the BH, the Bardeen metric is more similar to the Schwarzschild metric than the RN one, what is reflected in the emitted power. 

\begin{figure*}[!h]
	\includegraphics[scale=0.7]{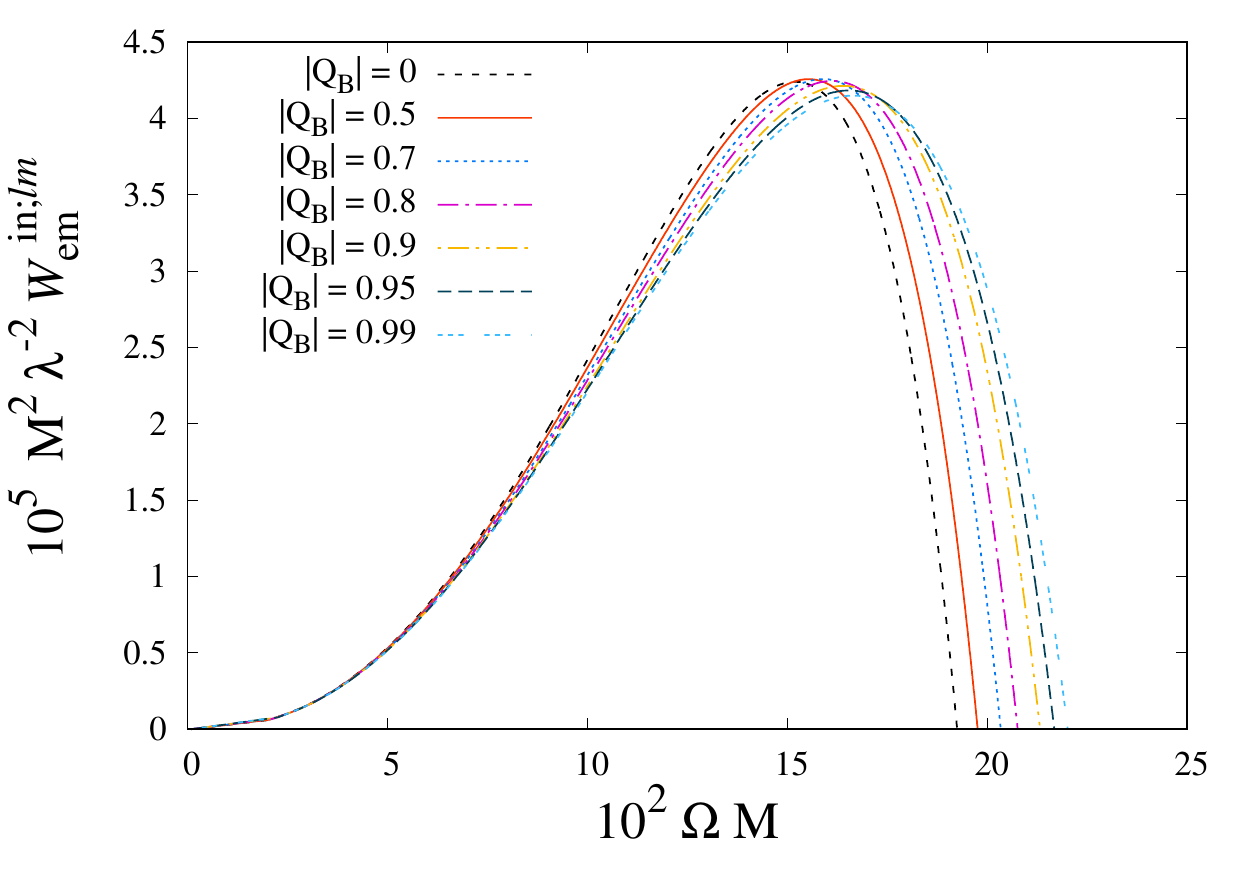} \includegraphics[scale=0.7]{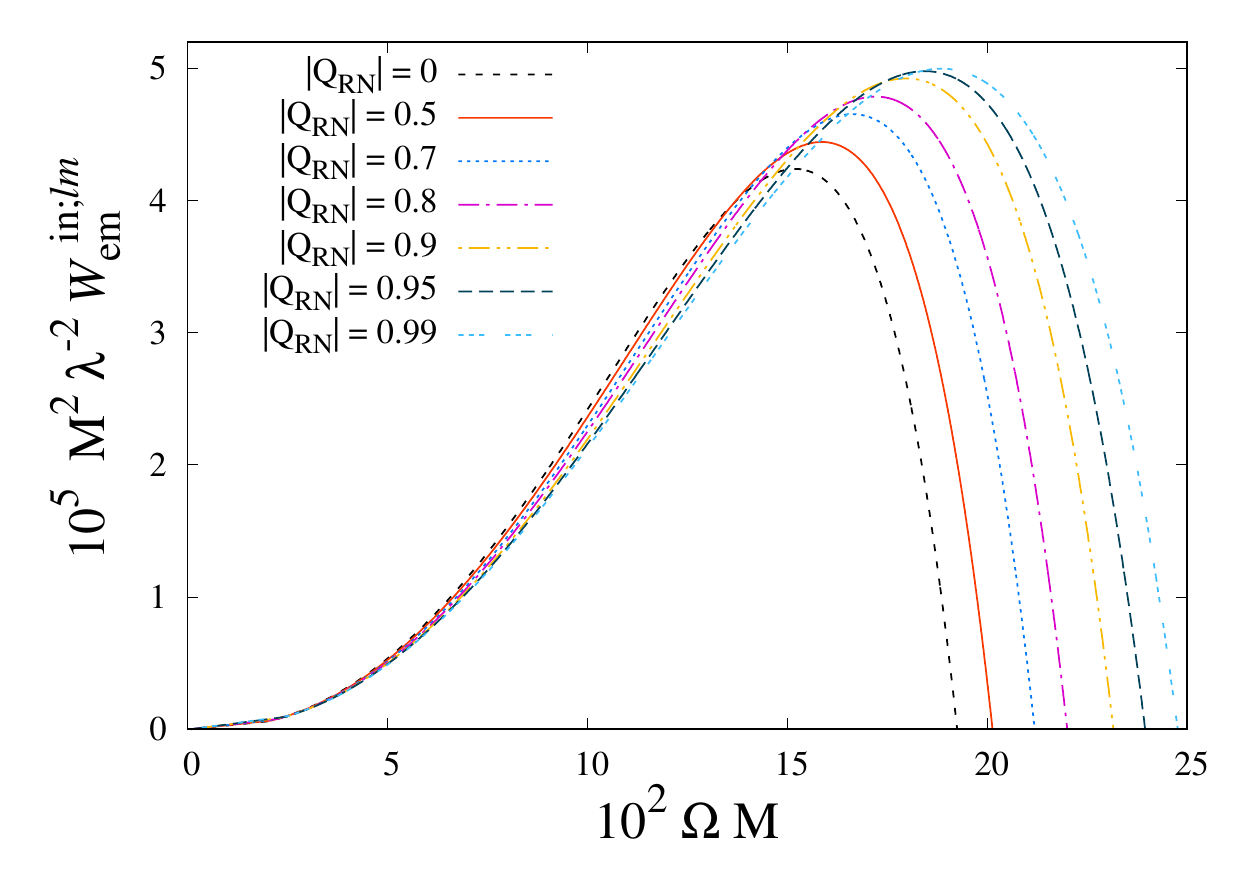}  \\
	\vspace{0.33cm}
	\includegraphics[scale=0.7]{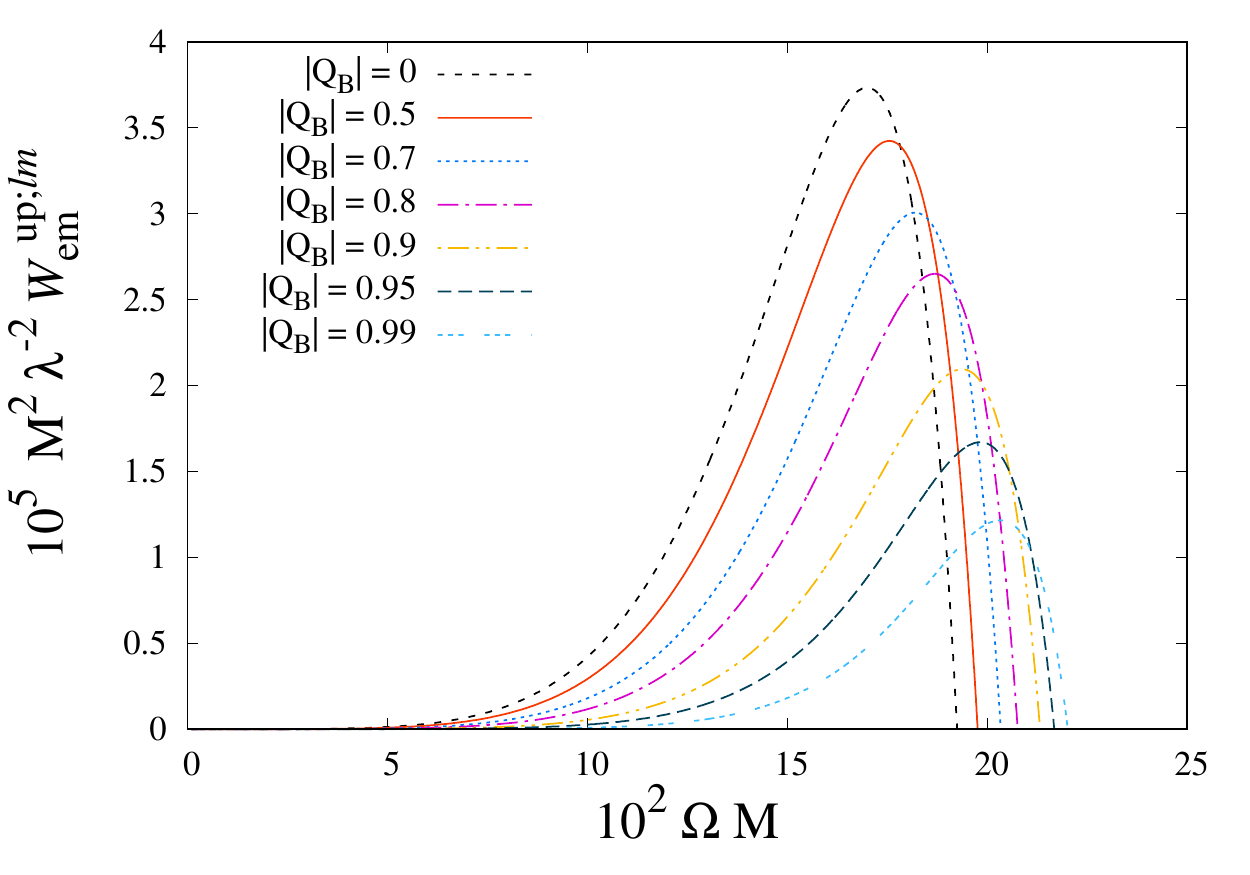} \includegraphics[scale=0.7]{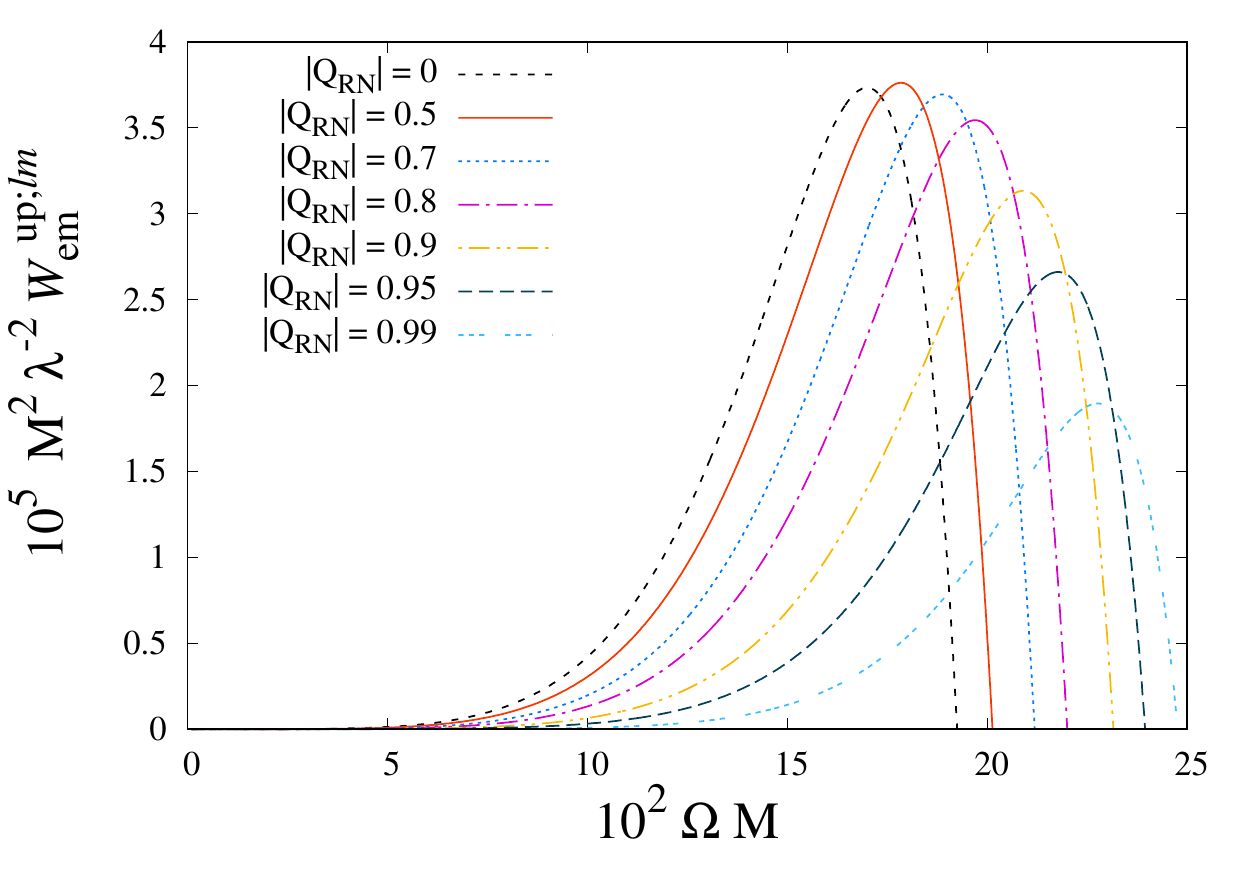}  \\
	\vspace{0.33cm}
	\includegraphics[scale=0.7]{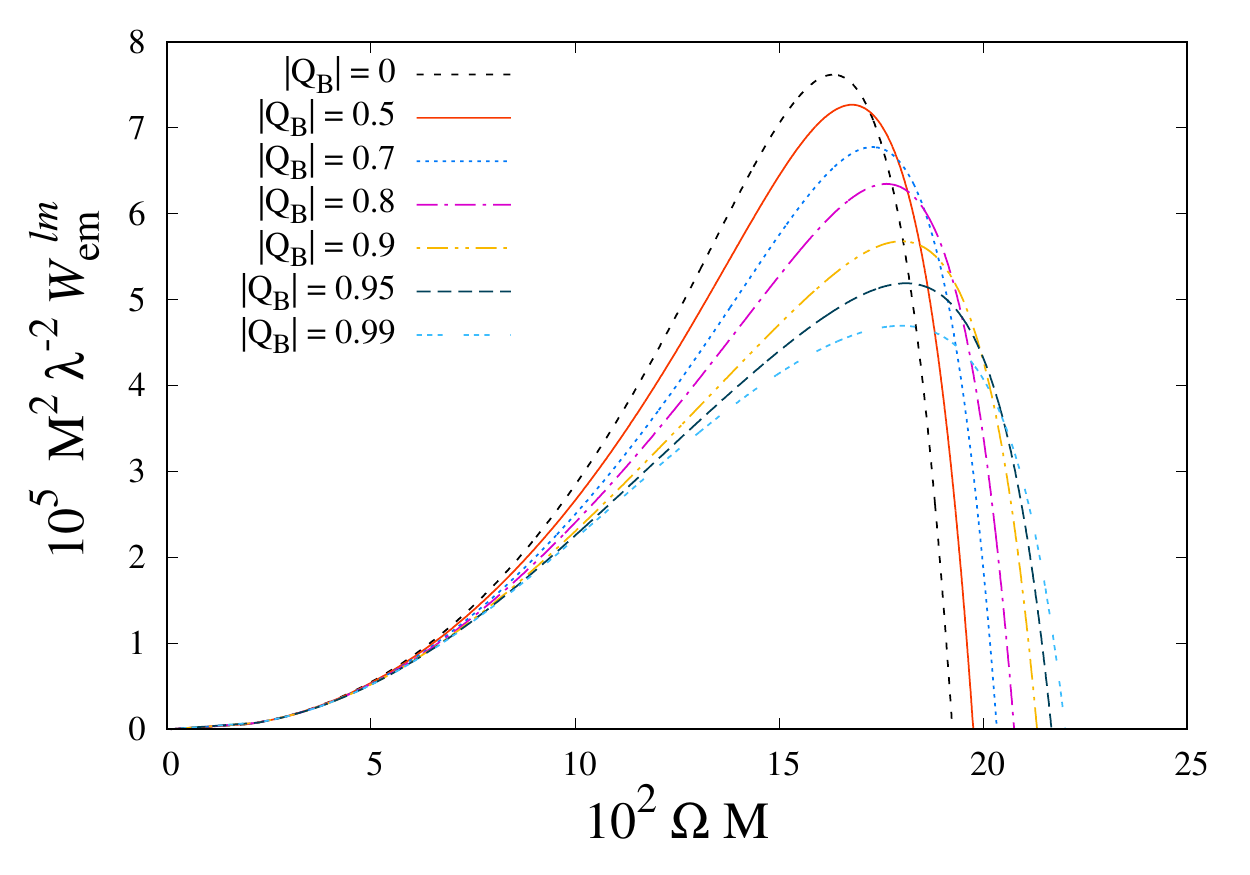} \includegraphics[scale=0.7]{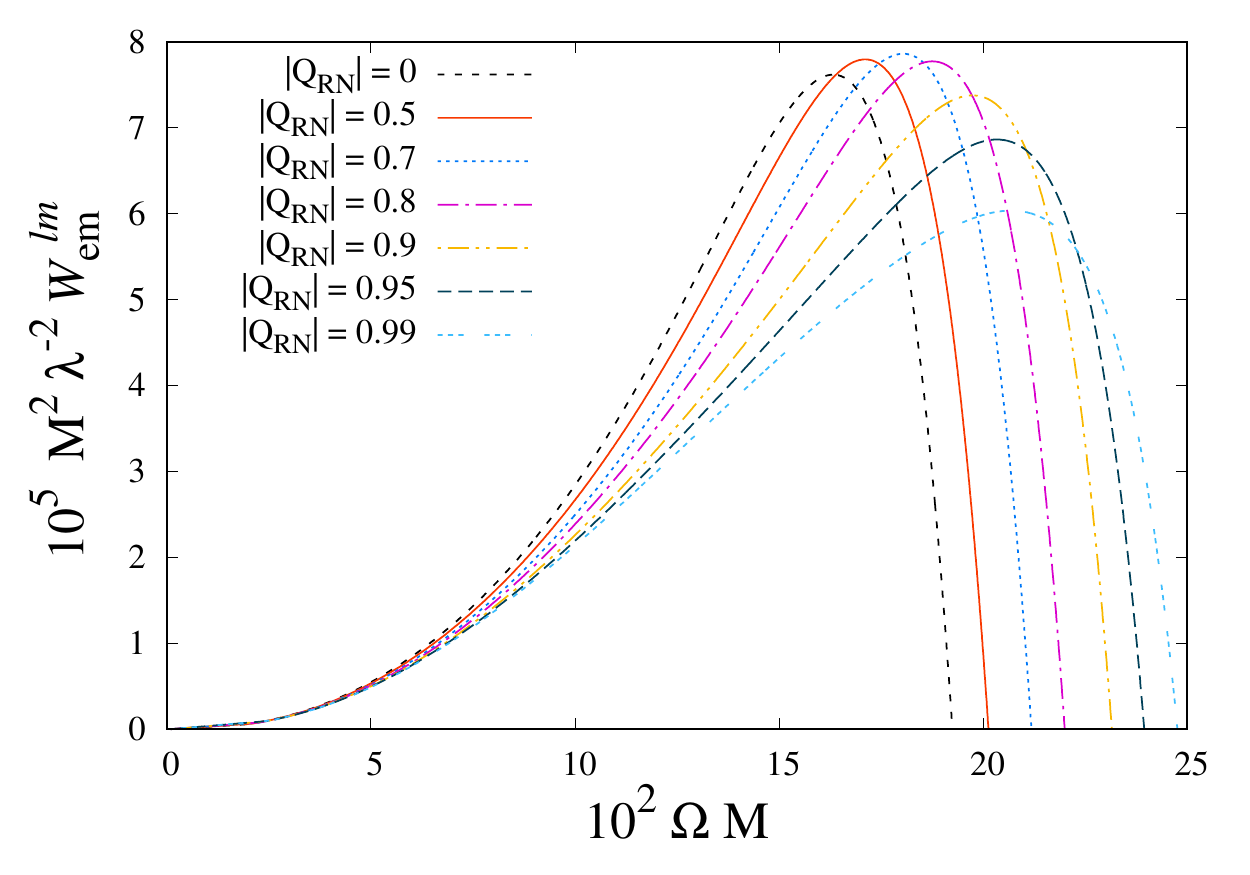}
	\caption{Emitted power for $in$ (\textit{top}), $up$ (\textit{middle}) modes and their sum (\textit{bottom}) of the $l=m=1$ mode, as a function of $\Omega$, in the Bardeen BH spacetime (\textit{left}), and in the RN BH spacetime (\textit{right}), with different values of $|Q_B|$ and $|Q_{RN}|$, respectively.}
	\label{fig:differentqs-total}
\end{figure*}

We also compute the emitted power for different values of the charge $Q$. We see, in Fig.~\ref{fig:differentqs-total}, that the peak of the total emitted power decreases as we increase the charge of the Bardeen BH. In the RN spacetime, the behavior is more involved. We first see an increase in the peak of the total emitted power as we increase the charge and, from a certain value of the charge on, the behavior is similar to the Bardeen BH case. The main difference between the Bardeen and RN cases is that the $in$ mode peaks have a notable increment as we increase the charge in the RN case (see the top plot in the right panel of Fig.~\ref{fig:differentqs-total}). Moreover, the \textit{up} mode peaks in the RN BH case have a slower fall off as we increase the charge, what can be seen in the middle right plot of Fig.~\ref{fig:differentqs-total}.
 
\begin{figure}[h!]
	\includegraphics[scale=0.7]{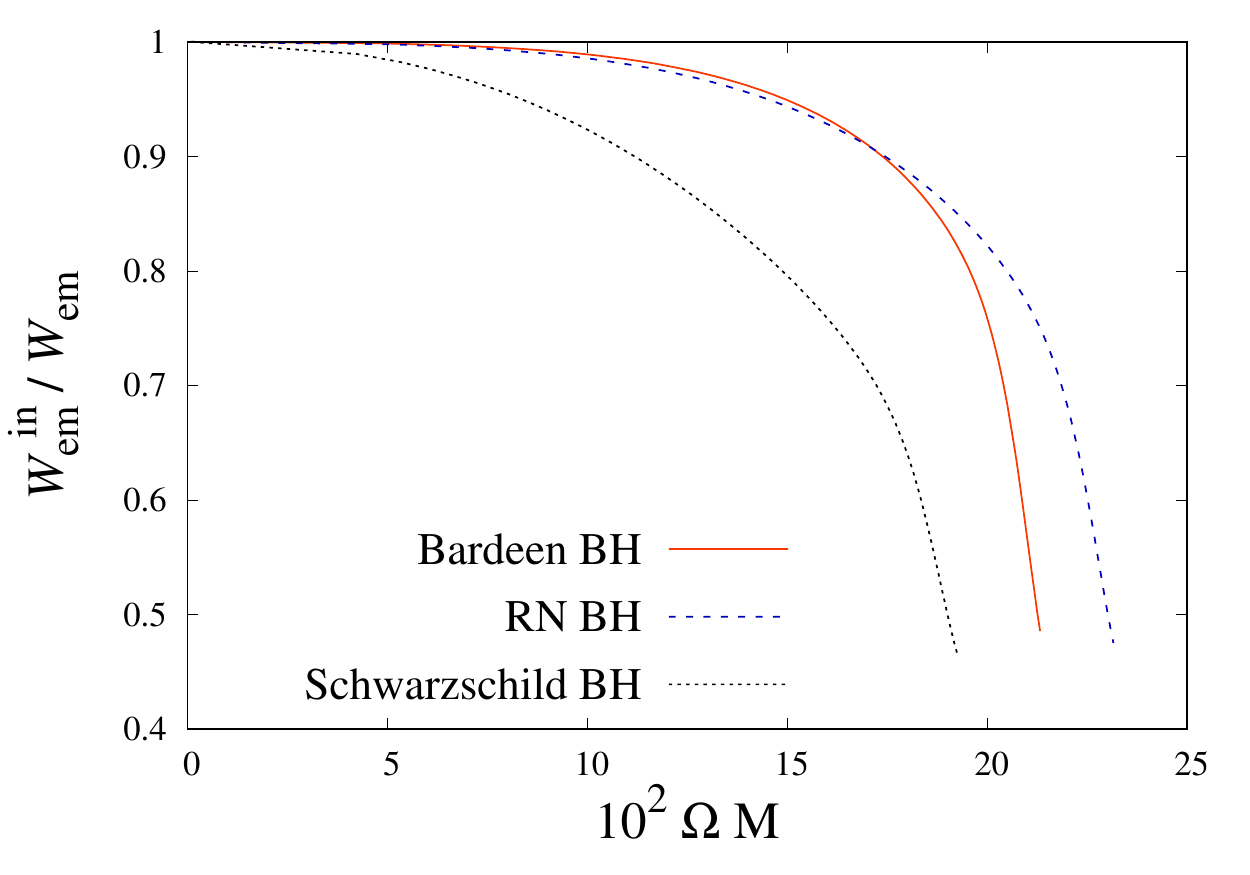}
	\caption{Ratio between the observed power at infinity, given by Eq.~(\ref{eq:observed-power}), and the total emitted power, given by Eq.~(\ref{eq:total-emitted-power}). The truncation in the $l$ summations of Eqs.~(\ref{eq:total-emitted-power}) and (\ref{eq:observed-power}) is done at $l_{\mathrm{max}}=20$. Both the Bardeen BH and the RN BH have been chosen such that $|Q_{(i)}|=0.9$.}
	\label{fig:ratio}
\end{figure}

The total emitted power is given by
\beq
W_{\mathrm{em}}=\sum_{P=in, up}\sum_{l,m}W^{P;lm}_{\mathrm{em}}. \label{eq:total-emitted-power}
\eeq
The observed power at infinity is obtained by summing only the contributions from the $in$ modes in Eq.~(\ref{eq:total-emitted-power}) (see, for example, Ref.~\cite{Crispino2016}), namely
\beq
W^{in}_{\mathrm{em}}=\sum_{l,m}W^{in;lm}_{\mathrm{em}}, \label{eq:observed-power}
\eeq
 which can be compared to the total emitted power. In Fig.~\ref{fig:ratio}, we plot the ratio between the observed power at infinity, $W^{in}_{\mathrm{em}}$, given by Eq.~(\ref{eq:observed-power}), and the total emitted power, given by Eq.~(\ref{eq:total-emitted-power}). For a Bardeen BH and a RN BH, this ratio behaves very similarly for low to intermediate values of the angular velocity (up to $\Omega M \approx 0.175$), with the radiation in the former case being slightly higher. For angular velocities higher than $\Omega M \approx 0.175$, the radiated power reaching null infinity in the RN case is higher than the one emitted in the Bardeen setting. Moreover, we see that, in both RN and Bardeen cases, the amount of radiation reaching infinity is higher in comparison with the Schwarzschild case, for the same value of the orbital angular velocity. 

\begin{figure}
	\includegraphics[scale=0.7]{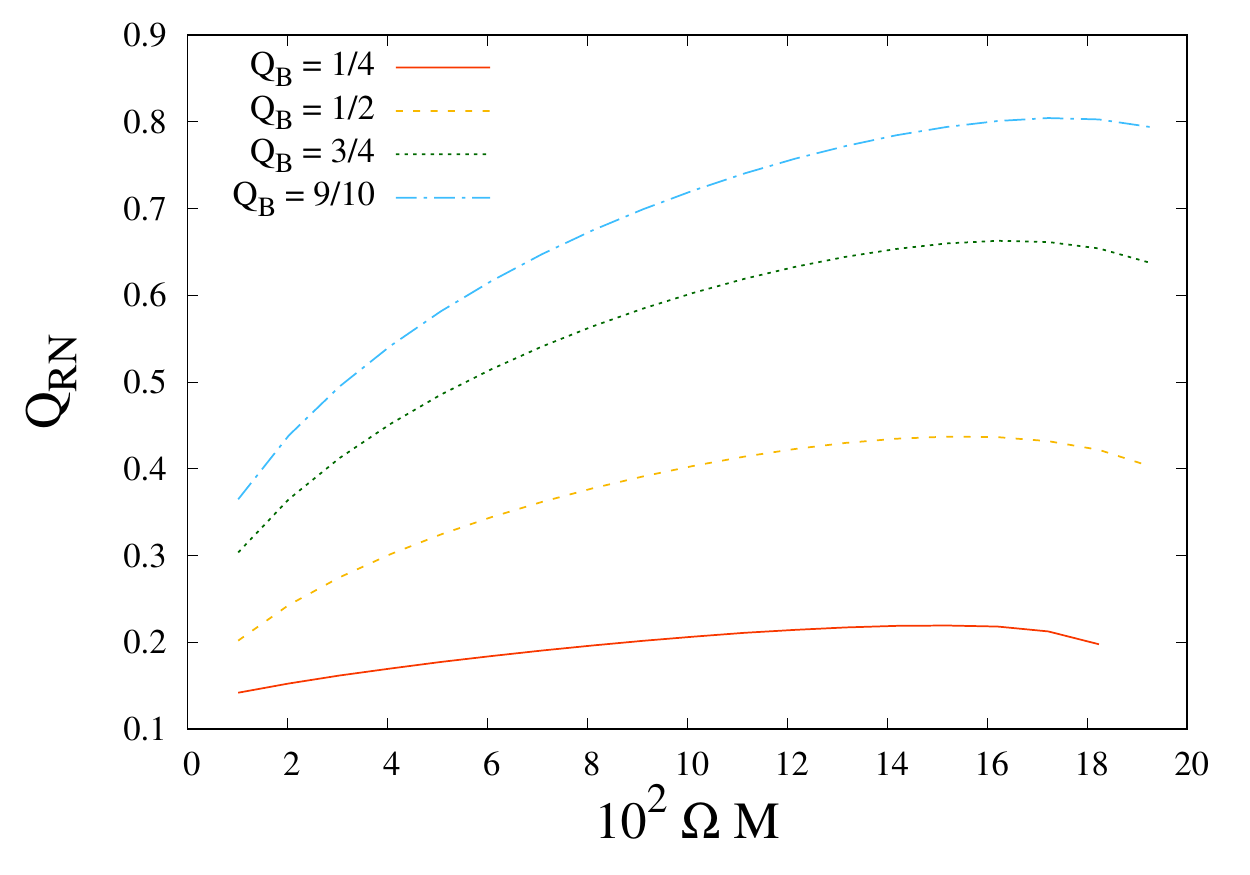}
	\caption{The equivalent normalized charge $Q_{RN}$ of a RN BH, in order for the total observed power at infinity to be the same as the one in the spacetime of a Bardeen BH with normalized charge $Q_{B}$.}
	\label{fig:compQ}
\end{figure}

For a given value of the source's angular velocity, the amount of total power radiated to infinity in the RN BH case can be the same as the one obtained in the Bardeen BH case, for specific values of the pair $(Q_B,Q_{RN})$. Given a value for the normalized charge $Q_B$ of a Bardeen BH, one can search for the corresponding normalized charge $Q_{RN}$ of a RN BH, such that the observed radiated power, given by Eq.~(\ref{eq:observed-power}), is the same in both cases. In Fig.~\ref{fig:compQ} we plot, for some representative values of $Q_{B}$, the normalized charge $Q_{RN}$ as a function of the source's angular velocity, up to the maximum allowed value in Schwarzschild spacetime ($\Omega M=\frac{1}{3\sqrt{3}}$). The $l$ summation in Eq.~(\ref{eq:observed-power}) was truncated at a certain value $l_{max}$, such that the computed observed powers, using truncations at $l_{max}$ and $l_{max}+1$, differ in less than $1\%$. As we increase the angular velocity, higher multipole modes start to significantly contribute to the total emitted power (synchrotron radiation). Consequently, the truncation in $l$ needs to be much higher than considered here and the convergence of our numerical results is affected. From Fig.~\ref{fig:compQ} we see that there is a considerable difference between the normalized charges in each case.

%\includegraphics[scale=0.6]{RN_Wobs_Q_09_l_1-5.pdf}

%\includegraphics[scale=0.6]{RN_Wobs_Q_09_l_10-30.pdf}

%\includegraphics[scale=0.6]{RN_W_Q_09_l_10-30.pdf}
%%%%%%%%%%%%%%%%%%%%%%%%%%%%%%%%%%%%%%%%%%%%%%
\section{Final Remarks}
\label{sec:finalremarks}
%%%%%%%%%%%%%%%%%%%%%%%%%%%%%%%%%%%%%%%%%%%%%%

In this paper we have computed the scalar radiation emitted by a source in circular geodesic orbit around a Bardeen black hole, which constitutes a regular spacetime. Using the quantum field theory in curved spacetimes approach, we obtained the emitted power by computing the one-particle-emission amplitude when the scalar field is excited by the external source. We compared this emitted power with the one obtained when the scalar source orbits a Reissner-Nordstr\"{o}m black hole. We have found that, as we increase the normalized charge in both cases, the peak in the emitted power of the Bardeen black hole case suffers a notable drop. In the Reissner-Nordstr\"{o}m case, we first observe a small increase to the emitted power peak and, as we continue to increase the charge, there is a moderate decrease of the peak. The difference in the emitted power between the two cases is mainly due to the behavior of the $in$ modes, which are the ones that contribute to the observed power at infinity.

We have also shown that the Bardeen black hole setting allows, for low to intermediate values of the source's angular velocity, more of the emitted radiation to reach null infinity in comparison to the case of a Reissner-Nordstr\"{o}m black hole, although their behavior are quite similar in this range. Beyond a certain value of the angular velocity, the emitted radiation reaching null infinity becomes higher in the Reissner-Nordstr\"{o}m black hole setting. In both cases, the observed radiation is higher than in the similar setting of a Schwarzschild black hole.  

The total radiated power observed at infinity in the Reissner-Nordstr\"{o}m black hole case can be equal to the same quantity in the Bardeen black hole setting, for appropriate values of the corresponding normalized charges. For a given value of the normalized charge $Q_B$ in the Bardeen case, our results show that the normalized charge $Q_{RN}$ of the Reissner-Nordstr\"{o}m black hole, needed for the equality between the total observed powers, is smaller than $Q_{B}$, for values of the source's angular velocity smaller than the maximum allowed value in Schwarzschild spacetime ($\Omega M=\frac{1}{3\sqrt{3}}$).

The radiation setting considered here also serves as a first step in considering more realistic scenarios such as the emission of electromagnetic and gravitational radiations in the spacetime of regular black holes, such as the Bardeen solution.

%%%%%%%%%%%%%%%%%%%%%%%%%%%%%%%%%%%%%%%%%%%%%%
\begin{acknowledgments}
We would like to thank A. Higuchi and C. Macedo for useful discussions. This research was financed in part by Coordena\c{c}\~ao de Aperfei\c{c}oamento de Pessoal de N\'ivel Superior (CAPES, Brazil) -- Finance Code 001, and by Conselho Nacional de Desenvolvimento Cient\'ifico e Tecnol\'ogico (CNPq, Brazil). This research has also received funding from the European Union's Horizon 2020 research and innovation programme under the H2020-MSCA-RISE-2017 Grant No. FunFiCO-777740.

\end{acknowledgments}
%%%%%%%%%%%%%%%%%%%%%%%%%%%%%%%%%%%%%%%%%%%%%%

%%%%%%%%%%%%%%%%%%%%%%%%%%%%%%%%%%%%%%%%%%%%%%
% Bibliography
%%%%%%%%%%%%%%%%%%%%%%%%%%%%%%%%%%%%%%%%%%%%%%

%\bibliographystyle{h-physrev4}
%\bibliographystyle{myutphys}
%\bibliographystyle{customapsrev4-1}
%\bibliography{references}


\end{document}